\documentclass[%
 aip,
 amsmath,amssymb,
 reprint,%
]{revtex4-1}

\usepackage{graphicx}
\usepackage{dcolumn}
\usepackage{bm}

\usepackage[utf8]{inputenc}
\usepackage[T1]{fontenc}
\usepackage{mathptmx}

\usepackage{color,xcolor}

\usepackage{array}
\newcolumntype{L}{>{\centering\arraybackslash}m{2.5cm}}

\usepackage[most]{tcolorbox}
\newtcolorbox{highlighted}{colback=yellow,coltext=red,breakable}

\newcommand{\ET}{\mathbf{E}_{\mathrm{THz}}}
\newcommand{\Eem}{\mathbf{E}_{\mathrm{em}}}
\newcommand{\Eout}{\mathbf{E}_{\mathrm{out}}}
\newcommand{\Pmat}{\mathbf{P}}
\newcommand{\Rmat}{\mathbf{R}}

\begin{document}


\title[Mosley \emph{et al.}]{Precise and accurate control of the ellipticity of THz radiation using a photoconductive pixel array}

\author{C.\ D.\ W.\ Mosley}%
\affiliation{University of Warwick, Department of Physics, Gibbet Hill Road, Coventry, CV4 7AL, United Kingdom.}
\affiliation{Lancaster University, Department of Physics, Lancaster, LA1 4YW, United Kingdom.}

\author{J. Deveikis}
\affiliation{University of Warwick, Department of Physics, Gibbet Hill Road, Coventry, CV4 7AL, United Kingdom.}

\author{James Lloyd-Hughes}
\email{j.lloyd-hughes@warwick.ac.uk}
\affiliation{University of Warwick, Department of Physics, Gibbet Hill Road, Coventry, CV4 7AL, United Kingdom.}

\date{\today}

\begin{abstract}
Full control of the ellipticity of broadband pulses of THz radiation, from linear to left- or right-handed circular polarization, was demonstrated via a 4-pixel photoconductive emitter with an integrated achromatic waveplate. 
Excellent polarization purity and accuracy was achieved, with Stokes parameters exceeding 97\,\% for linear and circular polarization, via a robust scheme that corrected electrically for polarization changes caused by imperfect optical elements.
Further, to assess the speed and precision of measurements of the THz polarization we introduced a new figure of merit, the standard error after one second of measurement, found to be $0.047^{\circ}$ for the polarization angle.
\end{abstract}

\maketitle


The accurate generation and precise detection of broadband THz beams with well-defined polarization states is of great importance for the spectroscopy of anisotropic materials such as birefringent materials\cite{Nagashima2013,Lloyd-Hughes2014,Mosley2017}, multiferroics\cite{Mosley2017} and quantum Hall systems.\cite{Failla2016} Polarization control is essential for the development of THz ellipsometry and polarimetry systems\cite{Nagashima2013,Watanabe2018} and for the nascent field of THz communications.
This has driven increased interest in polarization-control schemes that can modify the ellipticity and polarization angle of THz radiation. 

The direct generation of elliptical THz polarization states has been achieved by: photoconductive emitters under a magnetic field;\cite{Johnston02-127,Castro-Camus12-3620} superimposing multiple time-delayed THz pulses with different polarization states, as demonstrated for optical rectification,\cite{Amer2005} lateral photo-Dember\cite{Lee2012} and spintronic emitters;\cite{Chen2019} pulse-shaping of the pump pulse;\cite{Sato2013} laser-induced plasma filaments under an external electric field\cite{Lu2012} or two-color mixing of the pump.\cite{Zhang2018}
Alternatively, linearly-polarized THz can be converted to elliptical or circular polarization using broadband quarter waveplates based on birefringent materials\cite{Masson2006,Nagashima2013} or total internal reflection (TIR).\cite{Hirota2006,Kawada2014} 

In all these implementations the THz pulse polarization had a fixed ellipticity that was not circular at all frequencies, or the ellipticity could only be modified by the mechanical rotation or translation of a component, which is a slow and cumbersome process.
Full electrical control of the ellipticity would provide a route to accurate and precise THz polarimetry, imaging and ellipsometry.
In that regard a four-electrode photoconductive antenna has been shown to generate broadband left- or right-handed THz radiation via TIR, but the ellipticity angle varied from $35^{\circ}$ (elliptical) to $45^{\circ}$ (circular) over the experimental bandwidth, attributed to the inhomogeneous bias field of the stripline antenna.\cite{Hirota2006}

In this Article we report that multi-pixel photoconductive arrays can accurately control the polarization state of THz radiation from linear to circular. We further introduce quantitative measures that allow the accuracy and precision of polarization control to be established.
Our experimental approach used an array of interdigitated photoconductive antenna integrated with an achromatic quarter waveplate, and allows rapid electrical control of the polarization state between pure linear, left- or right-handed circular polarization, and in a way that corrected for the finite polarization performance of the optical setup. 

Interdigitated photoconductive emitters\cite{Dreyhaupt2005,Beck10-2611,Singh2019} are based on interleaved electrodes on the surface of a semiconductor that form an anode/semiconductor/cathode/gap repeat unit. 
Charge carriers are excited in the biased semiconductor by a fs optical pulse, creating a transient current that radiates a THz pulse, and THz pulses from each repeat unit interfere constructively in the far-field.
Interdigitated emitters benefit from lower drive voltages and superior radiation patterns than narrow-gap stripline or bowtie antennae, which exhibit rapidly diverging beams and require integrated Si correction lenses.
Further, their linear polarization purity is excellent, with an ellipticity ($< 1^{\circ}$) that is an order of magnitude better than that for bowtie and wide-gap THz emitters.\cite{Mosley2017} 


Here, a multi-pixel emitter array was produced by UV photolithography and consisted of four separate pixels, each consisting of interdigitated metal contacts with a 150\,$\mu$m $\times$ 150\,$\mu$m area, on a semi-insulating GaAs substrate. 
Adjacent pixels emitted orthogonally polarized THz polarization states, as defined by the direction of the applied electric field. 
The THz radiation emitted by the four pixels overlaps in the far-field to produce a coherent beam of THz radiation, with a linear polarisation state and a polarisation angle controlled by the relative bias voltages applied to horizontally-emitting and vertically-emitting pixels.\cite{Mosley2019a,Maussang2019}
The emitter was photoexcited by 80\,fs pulses from a Ti:sapphire oscillator, with an average optical power of 350\,mW, and a beam larger than the entire 300\,$\mu$m $\times$ 300\,$\mu$m active area of the device.

\begin{figure}
\includegraphics[width=0.5\textwidth]{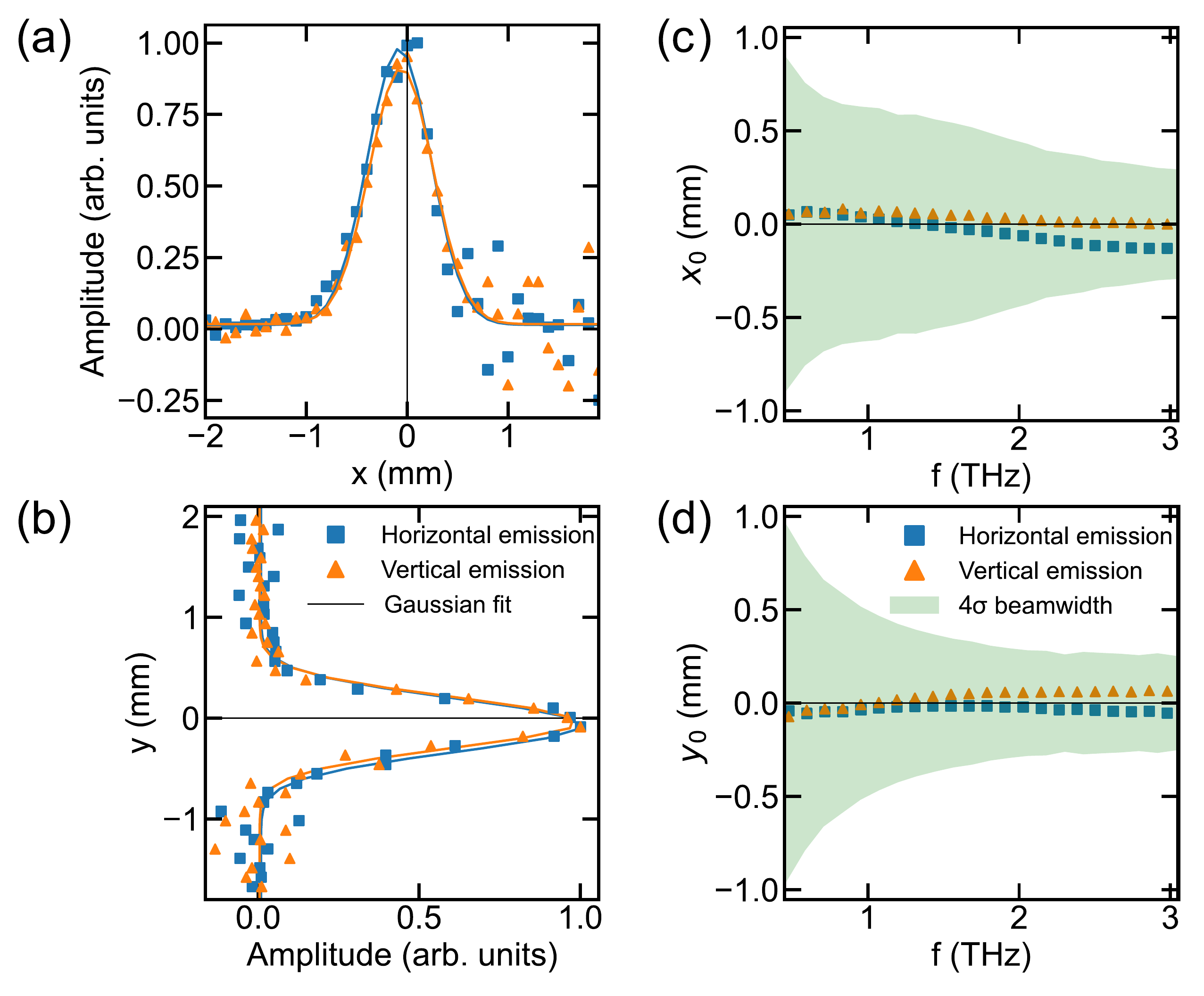}%
\caption{\label{Figure1} Measured beam profiles at 1\,THz for horizontally and vertically emitting pixels, scanning (a) horizontally, along $x$, and (b) vertically, along $y$. (c) and (d) Frequency dependence of beam width and beam center position for horizontally and vertically polarized emission, along horizontal and vertical directions respectively.}%
\end{figure}

The accurate control of the polarization state is reliant upon achieving good spatial overlap between THz beams produced from different pixels. To demonstrate that this is the case for multi-pixel emitters, we determined the frequency-dependent beam profile at the THz beam focus produced by a 3" focal length, 2" diameter off-axis parabolic mirror.
The THz time-domain waveform was measured at each position of a razor blade, which was stepped either horizontally or vertically through the focus. Both the horizontal, $E_x$, and vertical, $E_y$, components of the electric field of the THz pulse were detected using polarization-resolved electro-optic sampling\cite{VanderValk2005} in a 0.5\,mm-thick, [111]-oriented ZnTe crystal. The beam profiles, obtained by differentiating the spectral amplitude at a particular frequency versus position, are reported as points in Fig.\ \ref{Figure1}(a) and (b) for 1\,THz.
A Gaussian beam shape (solid lines) with centre positions $x_0$ and $y_0$ and standard deviation $\sigma_x$ and $\sigma_y$ was obtained at all frequencies, with frequency-dependent beam parameters reported in panels (c) and (d). The beam width, shown as $\pm2\sigma$, reduces at higher frequencies, as expected from Gaussian beam theory. 
Importantly, $\sigma_x\simeq\sigma_y$ for all frequencies, while $x_0$ and $y_0$ vary by less than 150\,$\mu$m, lower than the beam diameter (D4$\sigma$, given by the vertical extent of the shaded region). These measurements demonstrate that multi-pixel emitters produce high quality Gaussian beams, free from interference effects, and validate the beam propagation assumptions of our ellipticity-control setup, discussed in the following.

\begin{figure}
\includegraphics[width=0.95\columnwidth]{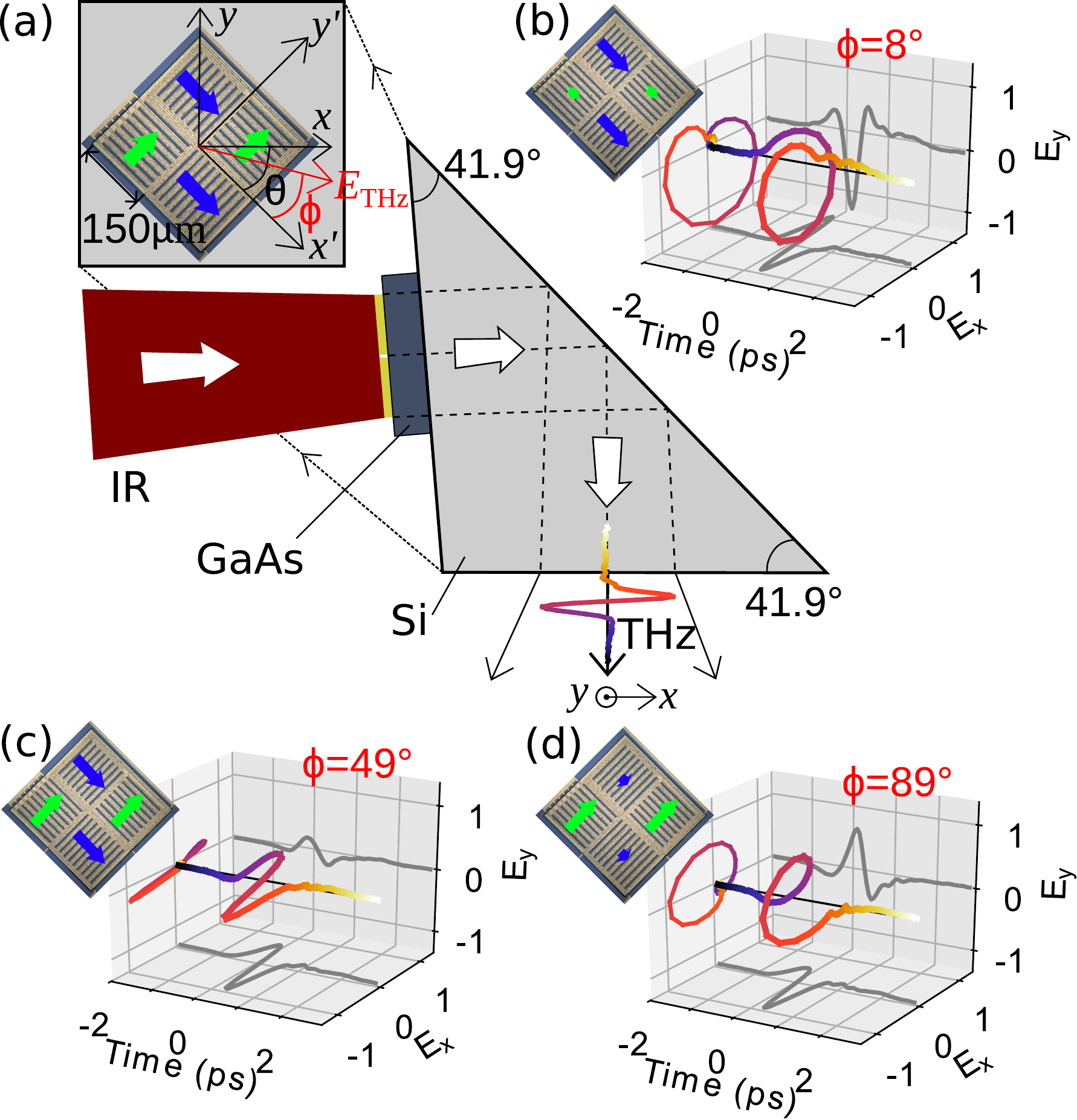}%
\caption{\label{Figure2} (a) Method to generate THz pulses with controllable ellipticity, as described in the text, based on using a 4-pixel THz emitter to produce a THz electric field pulse $\ET$ at a polarization angle $\phi$. Polarization-resolved THz time-domain waveforms measured for (b) circular right, (c) circular left and (d) linear horizontal THz radiation output, obtained using bias schemes (blue and green arrows) with different $\phi$.}%
\end{figure}

To demonstrate the electrical control of the THz polarization state from linear to circular we used the experimental setup presented in Fig.\ \ref{Figure2}(a).
The 4-pixel photoconductive emitter was optically contacted to one face of a Fresnel prism made of high-resistivity float-zone silicon, with its axes $x'$ and $y'$ at an angle $\theta=45$\,$^{\circ}$ to the lab frame, defined by $x$ and $y$. 
Square wave voltages with amplitudes $V_{H}=V_0 \cos\phi$ and $V_{V}=V_0\sin\phi$ were applied to pixels emitting $E_{x'}$ and $E_{y'}$, respectively, where $V_0=10$\,V and $\phi$ is the target emission angle in the emitter frame.
Thus, $\phi=0^{\circ}$ set an electric field bias across just two pixels (blue arrows) and created a THz electric field vector $\Eem=(E_{x'}, E_{y'})=(E_0, 0)$ in the emitter frame, and equal s- and p- components incident onto the silicon/air interface. 
At an angle of incidence close to 41.9$^{\circ}$, total internal reflection in the prism produced a phase advance $\delta=\delta_\mathrm{p}-\delta_\mathrm{s}=90^{\circ}$ for p-polarized THz (phase $\delta_p$) with respect to s-polarized THz (phase $\delta_s$), i.e.\ the prism acted as an achromatic quarter waveplate.\cite{Hirota2006} 
We reduced the divergence of the THz beam within the emitter and prism by adopting a weakly-focusing IR excitation beam, which transferred the pulse wavefront of the IR beam onto the THz beam\cite{Beck10-2611} and prevented a large spread in the angle of incidence at the silicon/air interface, which would have altered $\delta$ away from the desired value. 
After TIR, a circularly polarized pulse therefore resulted. 
Alternatively, driving all pixels with the same bias (green and blue arrows, $\phi=45^{\circ}$) produced $(E_{x'}, E_{y'})=(E_0, E_0)$, and linearly p-polarised light before and after TIR.

Polarization- and time-resolved waveforms of the THz electric field pulses, detected after focusing the emitted THz onto the electro-optic crystal, are reported in Fig.\ \ref{Figure2}(b)-(d) for the different bias schemes pictured. When the emission angle was close to $x'$ or $y'$ in the emitter frame (i.e.\ $\phi\simeq 0^{\circ}$ and $\phi\simeq 90^{\circ}$) the THz radiation emitted was close to circular, while linearly polarized THz resulted when $\phi\simeq 45^{\circ}$.

\begin{figure}
\includegraphics[width=0.5\textwidth]{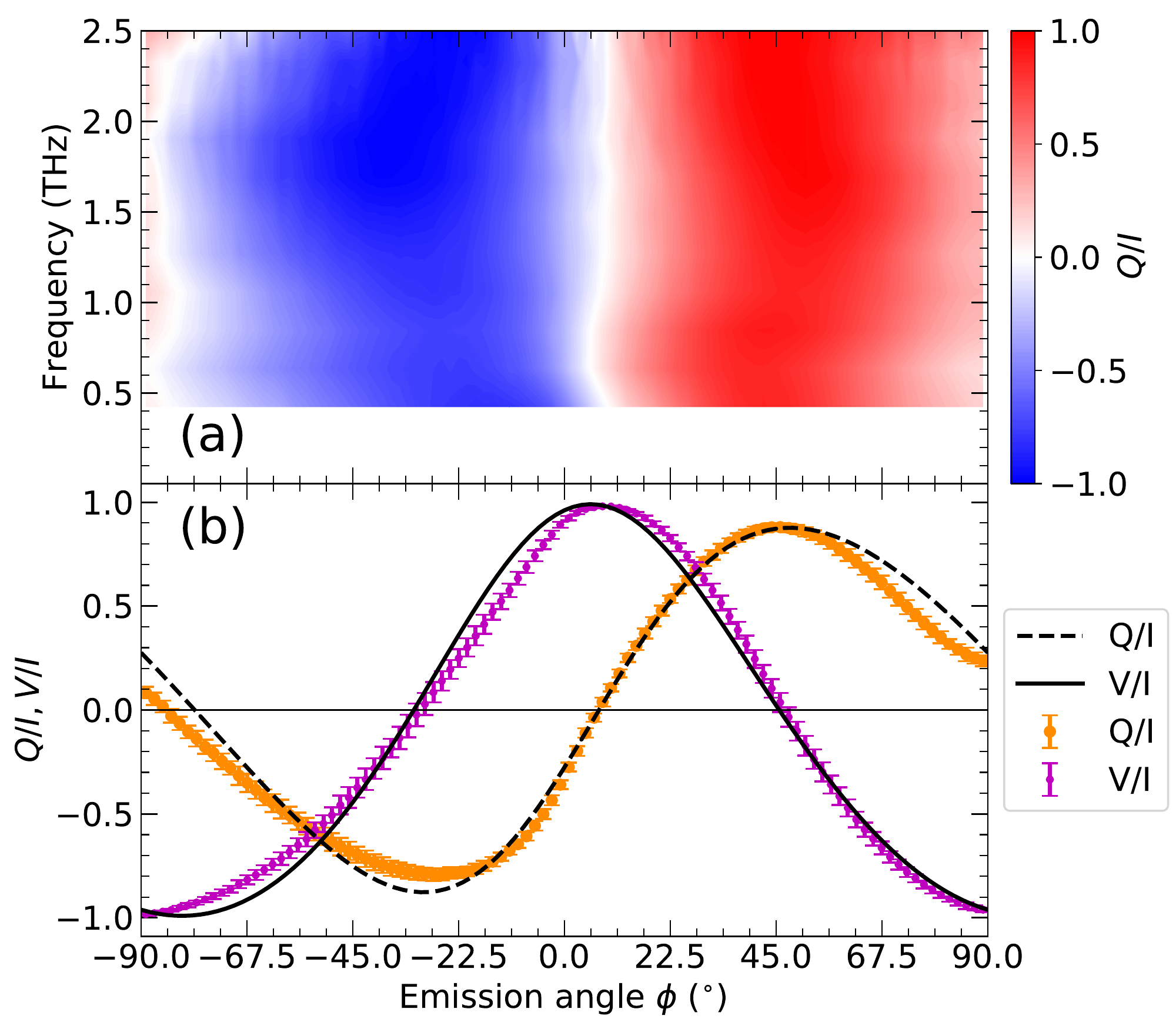}%
\caption{\label{Figure3}(a) Dependence of the normalized Stokes parameter $Q(\omega)/I(\omega)$ on frequency $\omega/2\pi$ and emission angle.
(b) Normalized Stokes parameters of the detected THz pulse averaged over the experimental bandwidth, from experiment (points) and model (lines).}%
\end{figure}

The accurate control of the ellipticity from this emitter design was demonstrated by a set of experiments at different $\phi$, and which are summarized in Fig.\ \ref{Figure3}. 
Polarisation-resolved time-domain THz waveforms were recorded for each $\phi$. 
The Stokes parameters, $\mathbf{S}(\omega,\phi)=(I,Q,U,V)$, provide a convenient figure of merit by which the accuracy of setting a desired polarization state can be assessed, where $I=|E_{x}|^{2}+|E_{y}|^{2}$, $Q=|E_{x}|^{2}-|E_{y}|^{2}$ is the difference in intensities polarised along $x$ and $y$, $U=|E_{a}|^{2}-|E_{b}|^{2}$ is the difference for light at $\pm45^{\circ}$ to the $x$ direction and $V=|E_{r}|^{2}-|E_{l}|^{2}$ is the difference for right- and left-hand circularly polarized light.
%
%

The frequency-dependent Stokes parameters were readily extracted from the Fourier transform of the time-domain traces to get $E_{x,y}(\omega)$, or $E_{a,b}(\omega)$ and $E_{r,l}(\omega)$ after converting into a $45^{\circ}$ or circular basis. The normalized Stokes parameter $Q/I$ is presented for the experimental frequency bandwidth in Fig.\ \ref{Figure3}(a), while panel (b) illustrates the frequency-averaged values of $Q/I$ and $V/I$. The electrical control of the ellipticity can be readily seen: upon changing the bias voltage applied to the horizontally and vertically emitting pixels, the polarization changes from circular ($V/I=\pm1$ and $Q/I=0$) to linear ($V/I=0$ and $Q/I=\pm1$), moving through elliptical polarisation states in-between (where $V/I$ and $Q/I$ are both nonzero). At emission angles of $8^{\circ}$ and $-89^{\circ}$ the value of $V/I\approx\pm1$ and the value of $Q/I$ drops to zero, demonstrating that the THz pulse after the prism was almost purely circularly polarised.

The small angular offset from where only the horizontally- or vertically-emitting contacts were biased, at $0^{\circ}$ and $\pm90^{\circ}$ respectively, can be attributed to a small angular misalignment of the emitter away from $45^{\circ}$ during mounting onto the prism, and small differences in emission strength or phase from different pixels. A distinct advantage of this method is that the user can electrically compensate for these issues, simply by varying the voltages $V_H$ and $V_V$ applied. The scheme can be summarized mathematically via 
\begin{equation*}
\Eout=\Pmat\,\Rmat\,\Eem = \begin{pmatrix}
1 & 0 \\
0 & i 
\end{pmatrix}
\begin{pmatrix}
\cos\theta & \sin\theta \\
-\sin\theta & \cos\theta
\end{pmatrix}
\begin{pmatrix}
E_0\cos\phi  \\
\alpha E_0 \sin\phi e^{i\beta} 
\end{pmatrix}
\end{equation*}

\noindent where $\Pmat$ is the Jones matrix for the waveplate, $\Rmat(\theta)$ is the rotation matrix, $\theta=\pi/4 + \epsilon$ to account for a small misalignment of the emitter by angle $\epsilon$ relative to $\theta=45^{\circ}$, and $\alpha$ and $\beta$ account for variations in the relative amplitude and phase of THz radiation emitted along $E_{x'}$ and $E_{y'}$, or resulting from optical components in the beam path.
This expression was used to calculate the $x$ and $y$ components of $\Eout$, and thus the Stokes parameters using their definitions above. Good accord with experiment was found using $\alpha=1.25$, $\beta=30^{\circ}$ and $\epsilon=8^{\circ}$, as reported in Fig.\ \ref{Figure3}(b).

Further insight into how this experimental scheme achieves polarization control can be obtained by considering $\Eout$ near $\phi=0$ and for small $\beta$ and $\epsilon$, which approximates to $\Eout=(E_l,E_r)=E_0(-\epsilon + \alpha\phi,1)$ in the circular basis. 
Hence by setting $\phi=\epsilon/\alpha\simeq 6^{\circ}$ one can produce right-handed circularly polarized light, as seen by $V/I=+1$ in Fig.\ \ref{Figure3}(b). 
Similarly, at $\phi=\alpha\epsilon \pm\pi/2$, pure left-handed THz can be obtained, with $V/I=-1$.
Hence, to first order this experimental scheme controls the ellipticity by setting $\phi$ to compensate for the amplitude difference $\alpha$ between $E_{x'}$ and $E_{y'}$, and the angular alignment error $\epsilon$.
The phase difference $\beta$, which may be caused in part by the finite ellipticity of the multi-pixel design ($<15^{\circ}$),\cite{Mosley2019a} does not affect the ellipticity near $\phi=0$ or $\phi=\pm\pi/2$ to first order.

\begin{figure}
\includegraphics[width=0.5\textwidth]{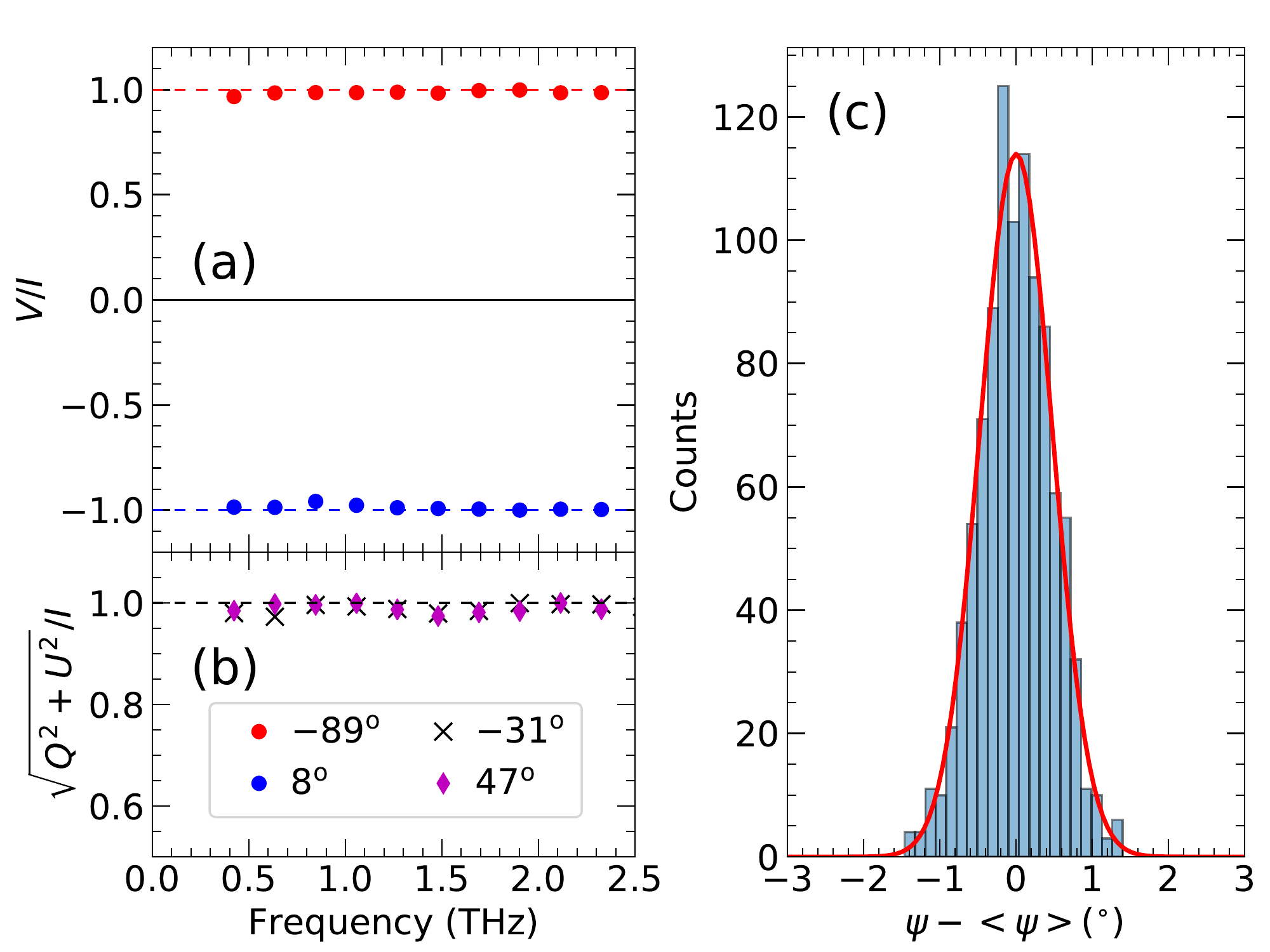}%
\caption{\label{Figure4} (a) Frequency dependence of Stokes parameters of the right-hand circularly polarised ($\phi=-89^{\circ}$) and left-hand circularly polarised ($\phi=8^{\circ}$) THz pulses, respectively. (b) Frequency dependence of linear polarisation purity for $\phi = -31^{\circ}$ and $\phi = 47^{\circ}$. (c) Histogram of the deviation from the mean, $\psi-<\psi>$, over 1000 repeated measurements of the polarization angle $\psi$, each taking 10\,ms. A Gaussian fit (red line) yields standard deviation $\sigma=0.47^{\circ}$ and $<\psi>=12.8^{\circ}$.}%
\end{figure}

To establish the performance of this system, and to enable comparison with other methods in the literature, we explored figures of merit associated with the accuracy and precision of the polarisation state generated and detected with this approach. The Stokes parameters provide a convenient figure of merit for the accuracy of the polarisation state, and are presented for the optimum circular and linear polarisation states in Figs.\ \ref{Figure4}(a) and \ref{Figure4}(b), respectively. Here we considered the combined contribution of both $Q$ and $U$ to the linear polarisation purity, using $\sqrt{Q^{2}+U^{2}}/I$, since a linear polarisation state at an angle away from the $x$ or $y$ axes has components $\pm45^{\circ}$ that contribute to $U$. High polarisation purities were achieved for both circular and linear polarizations, with $|V/I|>0.98$ and $\sqrt{Q^{2}+U^{2}}/I>0.97$ over the bandwidth of the experiment (2.5\,THz).

\begin{table*}[!t]
\centering
\begin{tabular}{|l|l|l|l|l|l|l|}
\hline
\textbf{Ref.}  & \textbf{Laser} & \textbf{Scheme}                                               & \textbf{$\sigma_{\psi} (^{\circ})$} & \textbf{$\tau$ (ms)} & \textbf{$R$ (Hz)} & \textbf{$s_{\psi}$ ($^{\circ}$)} \\ \hline
[\onlinecite{Yasumatsu2012}] & Amp.      & EOS with (110) GaP, continually rotating.                                            & 0.56                         & 21                   & 47.6              & 0.081                                       \\ \hline
[\onlinecite{Nemoto2014}]   & Amp.      & EOS with (111) ZnTe, gate polarization modulated via PEM.             & 0.0057                       & 660                  & 1.5               & 0.005                                       \\ \hline
[\onlinecite{Xu2020}]        & Amp.      & EOS with (110) ZnTe, continually rotating.                                           & 1.3                          & 62                   & 16                & 0.325                                       \\ \hline
[\onlinecite{Peng2020}]      & Osc.     & PCD with cross-polarized InP nanowires. & 0.38                         & 1000                 & 1                 & 0.38                                        \\ \hline
This work      & Osc.     & EOS with (111) ZnTe, gate polarization rotated via half-waveplate.     & 0.47                         & 10                   & 100               & 0.047                                       \\ \hline
\end{tabular}
\caption{\label{TAB:comparison}Comparison of calculated $s_{\psi}$ for papers that report $\sigma_{\psi}$ and $\tau$ or $R$. `Amp.' stands for Ti:sapphire laser amplifier; `Osc.' for Ti:sapphire laser oscillator; `EOS' denotes electro-optic sampling; and `PCD' refers to photoconductive detection.}
\end{table*}

Finally, we introduce the standard error in a measurement time of $T=1$\,s as a figure of merit that allows us to quantify the precision achieved in this work. 
The standard error of the mean is $s=\sigma/{\sqrt{N}}$, where $\sigma$ is the standard deviation of $N$ repeated measurements of an observable. 
Since $N=RT$ for a measurement rate $R$, $s=\sigma/{\sqrt{RT}}=\sigma \sqrt{\tau/T}$ where $\tau=1/R$ is the time to make one measurement.
By calculating $s$ for $T=1$\,s a fair comparison can be made of the precision of different schemes that measure the polarization of THz radiation, which have widely different sampling times, $\tau$ (Table \ref{TAB:comparison}).
As demonstrated above, the output ellipticity in our scheme depends on the set value of $\phi$, hence we measured the precision in the polarization angle $\psi$ produced by the 4-pixel emitter for a fixed value of $\phi$. 
Here we define $\psi$ as the measured angle of the THz pulse relative to the $x$ direction.
We determined $\sigma_{\psi} = 0.47^{\circ}$ from 1000 repeated measurements of $E_x$ and $E_y$ at a single point in the time-domain, as reported in Fig.\ \ref{Figure4}(c)). 
The sampling time, $\tau=10$\,ms, in this case was limited by the time constant of the lock-in amplifier used to average the electro-optic signal, giving $N=100$ in 1\,s.
Thus $s_{\psi}=\sigma_{\psi}/\sqrt{100}=0.047^{\circ}$, which is competitive with the current state of the art. 
In Table \ref{TAB:comparison} we report our calculated values of $s_{\psi}$ for various polarization-resolved detection schemes reported in the literature. 
To further lower $s_{\psi}$ and enhance precision it is desirable to either lower $\sigma$ by increasing the THz signal-to-noise ratio, as shown for high-field THz pulses from LiNbO$_3$,\cite{Nemoto2014} or using schemes that increase $R$.

In conclusion, full electrical control of the ellipticity of broadband THz pulses was obtained using a 4-pixel interdigitated photoconductive antenna with an integrated achromatic waveplate. 
High polarisation purities over the entire experimental bandwidth (0.2-2.5\,THz) were obtained, for both linear ($>97\%$) and circular ($>98\%$) polarisation states. Here the bandwidth was limited by the fs laser's pulse duration and increasing absorption at high frequencies in the Si prism.
Our scheme has the distinct advantage of allowing the user to electrically compensate for small misalignments of optical components, or the effects of optical components on the amplitude and phase of the THz radiation, to produce more accurate and pure polarisation states.
The versatile electrical control of THz polarization states using multi-pixel photoconductive emitters will find widespread application in industrial and commercial THz time-domain spectrometers, the majority of which use laser oscillators. 

Finally, we introduced the standard error for a 1\,s measurement time as a convenient metric for the precision, facilitating a fair comparison with different schemes reported in the literature. 
This metric can be applied universally across any experimental observable and allows the experimenter to discern the fastest, most precise scheme.

\begin{acknowledgments}
CDWM would like to thank the EPSRC (UK) for a PhD studentship. The authors would like to thank Hugh Thomas and Lucas Bartol-Bibb for technical assistance.

\end{acknowledgments}


%


\end{document}